\input amstex
\magnification=\magstep 1
\hsize=140mm 
\vsize=202mm
\hoffset=-3mm 
\voffset=-4mm
\parindent=0mm \parskip=4pt plus1pt minus1pt 

\font\hbf=cmbx10 scaled\magstep2 
\font\bbf=cmbx10 scaled\magstep1 
\font\sbf=cmbx9
\font\trm=cmr7
\font\srm=cmr9 
\font\caps=cmcsc10 
\def\newline{\hfil\break\noindent}
\def\newpage{\vfill\eject\noindent}
\def\mgap{\par\vskip 3mm\noindent} 
\def\bgap{\par\vskip 5mm\noindent} 
\loadmsbm 
\newsymbol\diagup 231E
\newsymbol\diagdown 231F
\def\aa{{\Cal A}}   \def\cc{{\Cal C}}   \def\hh{{\Cal H}}  \def\kk{{\Cal K}}
\def\a{\alpha}      \def\b{\beta}       \def\sig{\sigma}   \def\rho{\varrho}
\def\eps{\varepsilon}     \def\arr{\!\rightarrow\!}   \def\inv{^{-1}}
\def\mod{\;\hbox{mod}\;}  \def\eu{\hbox{e}}           \def\SU{\hbox{SU(2)}}
\def\ppartial{\vbox{\hbox{$\leftrightarrow$}\vskip-7pt\hbox{$\,\partial$}}}
\def\eins{{\mathchoice {\hbox{\rm 1}\mskip-4mu\hbox{\rm l}} 
{\hbox{\rm 1}\mskip-4mu\hbox{\rm l}} {\hbox{\rm 1}\mskip-4.5mu\hbox{\rm l}} 
{\hbox{\rm 1}\mskip-5mu\hbox{\rm l}}}}
\def\CC{\Bbb C} \def\ZZ{\Bbb Z} 
\def\cross{\hbox{\vbox{\hbox{\hbox{$\nwarrow$}\kern-.4em\hbox{$\nearrow$}}%
\vskip-.65em\hbox{\hbox{$\swarrow$}\kern-.4em\hbox{$\searrow$}}}}}
\def\westcross{\vbox{\hbox{\hbox{$\nwarrow$}\kern-.4em\hbox{$\diagup$}}%
\vskip-.66em\hbox{\hbox{$\swarrow$}\kern-.4em\hbox{$\diagdown$}}}}
\def\eastcross{\vbox{\hbox{\hbox{$\diagdown$}\kern-.4em\hbox{$\nearrow$}}%
\vskip-.66em\hbox{\hbox{$\diagup$}\kern-.4em\hbox{$\searrow$}}}}
\def\vac#1{\langle\Omega,#1\Omega\rangle}
\def\text#1{\quad\hbox{#1}\quad}
\def\frac#1#2{{#1\over#2}}  
\def\Circ{\vcenter{\vbox{\hbox{$\scriptstyle\circ$}\hrule width.0ex height.45ex
                         \hbox{$\scriptstyle\circ$}}}}
\def\Double#1{\,\Circ #1 \Circ\,}   
\def\double#1{\,{}\colon #1 \colon\,}  
\def\trip{\vcenter{\vbox{\hbox{$.$}\hrule width.0ex height.45ex
                         \hbox{$.$}\hrule width.0ex height.45ex \hbox{$.$}}}}
\def\triple#1{\,\trip #1 \trip\,}  
\def\K{1} \def\ST{2} \def\W{3} \def\B{4} \def\RS{5} \def\BPZ{6} \def\GKO{7}
\def\BRSS{8} \def\KZ{9} \def\WZW{10} \def\F{11} \def\LS{12} \def\SS{13} 
\def\CRW{14} \def\M{15} \def\BB{16} \def\SSV{17} \def\FRS{18} \def\J{19} 
\def\VO{20} \def\KF{21} \def\Spt{22}
\def\cite#1{[#1]}
\def\today{\ifcase \month \or Jan \or Feb \or Mar \or Apr \or May \or
Jun \or Jul \or Aug \or Sep \or Oct \or Nov \or Dec \fi \space
\number\day, \number\year}
\noindent DESY 97-096 \hfill May 1997 
\mgap 
\centerline{\hbf A non-abelian square root} \vskip2mm
\centerline{\hbf of abelian vertex operators} 
\bgap
\centerline{{\caps K. Frieler} and {\caps K.-H. Rehren}}
\centerline{II.\ Institut f\"ur Theoretische Physik} 
\centerline{Universit\"at Hamburg ({\caps Germany})}
\bgap
\baselineskip11pt 
{\sbf Abstract.} 
{\srm Kadanoff's ``correlations along a line'' in the critical two-dimensional
Ising model \cite\K\ are reconsidered. They are the analytical aspect of 
a representation of abelian chiral vertex operators as quadratic polynomials,
in the sense of operator valued distributions, in non-abelian exchange fields.
This basic result has interesting applications to conformal coset models.
It also gives a new explanation for the remarkable relation between
the ``doubled'' critical Ising model and the free massless Dirac theory 
\cite\ST. As a consequence, analogous properties as for the Ising model 
order/disorder fields with respect both to doubling and to restriction along 
a line are established for the two-dimensional local fields with chiral SU(2) 
symmetry at level 2.}
\baselineskip13.8pt 
\vskip5mm {\bbf 1. Introduction and outline} \vskip3mm
Wightman distributions can be pointwise multiplied, unlike general 
distributions, and the product is again a Wightman distribution. This fact 
\cite{\W, Chap.\ V} is due to the spectral properties, i.e., support 
properties in momentum space, of Wightman fields. The new Wightman field 
resulting from this prescription is known as the $p$-product of the factor 
fields. 

The converse is in general not true: the square root of a Wightman distribution
is in general not a Wightman distribution. There is, however, at least one 
famous non-trivial example, due to Schroer and Truong \cite\ST. The 
exponentials of the
potential for the axial current of a free Dirac field in two dimensions can 
be defined (with an appropriate renormalization) and give rise to
local Wightman fields. The sine and the cosine, at a specific value of 
the coefficient in the exponential, turn out to be the $p$-squares of the 
order and disorder fields of the Ising model. This result establishes a
deep field theoretical relation between the ``doubled'' Ising model 
and the free Dirac theory. Analytically it means that the vacuum expectation 
values of the latter can be computed by taking square 
roots of v.e.v.'s of the former; the double-valuedness of the square root
gives rise to the mutual non-locality of the order and disorder fields.

The purpose of this paper is the discussion of another instance of nontrivial
$p$-product factorization and some of its consequences. The basic formula 
(3.4) represents a chiral abelian vertex operator (cf.\ Sect.\ 2.1) as a 
$p$-quadratic expression in terms of chiral non-abelian exchange fields 
(Sect.\ 2.2); the latter are the well-known chiral constituents of the
order and disorder fields of the critical Ising field theory \cite\RS.

The $p$-product of chiral abelian vertex operators is another abelian 
vertex operators, with a quadratic addition law for the charges.
Conversely, the $p$-factorization of abelian vertex operators
into abelian vertex operators of smaller charges is an almost trivial 
fact (cf.\ Sect.\ 2.1 below). In contrast, the new feature of our basic result
is the {\it non-abelian} nature of the constituents, reflected in their 
non-abelian braiding and fusion properties.

The $p$-product is well-defined also for non-local fields, as long as they 
satisfy the remaining Wightman axioms. We shall refer to such fields as 
non-local Wightman fields. Indeed, our basic result (3.4) is in terms of 
non-local (chiral) fields, while some of its applications pertain to local 
fields in two dimensions.

Viewing Wightman fields as operator valued distributions, the $p$-product
is the pointwise multiplication as a distribution, and the tensor product as 
an operator. The $p$-product, denoted by $\Phi\odot\Psi$, of two Wightman 
fields $\Phi$ and $\Psi$ on Hilbert spaces $H_\Phi$ and $H_\Psi$ therefore 
lives on $H_\Phi \otimes H_\Psi$ (its cyclic subspace may be smaller). 
Symbolically we write
$$ (\Phi \odot \Psi)(f) = \int d^dx f(x) \; \Phi(x) \otimes \Psi(x) . 
\eqno(1.1)$$ 
The precise statement is that the tensor product of Wightman distributions 
extends to singular smearing functions of the form $f(x) \delta^d(x-y)$ where 
$f$ is a smooth test function.

The first input for the basic result (cf.\ Sect.\ 3) is the old observation 
in the critical Ising model \cite\K\ due to Kadanoff (for the magnetization 
$\sig$ alone) and Kadanoff and Ceva (for mixed correlations involving also 
the disorder parameter $\mu$) that upon restriction along a line in 
two-dimensional space, the correlations of $\sig$ and $\mu$ satisfy certain 
``$\Gamma$-selection rules'' and simplify considerably. Their analytical form 
becomes that of the correlations of appropriately assigned electric charges 
in a one-dimensional Coulomb gas. 

Translated to the associated quantum field theory in 1+1 dimensions, this
observation is tant\-amount to a simple identification between the order and 
disorder fields along a time-like axis and chiral abelian vertex operators 
$E_\a$ of charge $\a=\pm\frac 12$ and scaling dimension $h=\frac 18$.
Explicitly, the Kadanoff-Ceva formulae read
$$ \qquad\left\{\quad \eqalign{\sig\vert_{x=0} & \cong E_{+\frac12}P_+ + 
E_{-\frac12}P_- \cr \mu\vert_{x=0} & \cong E_{-\frac12}P_+ + E_{+\frac12}P_- 
\cr } \right. \eqno(1.2a) $$
$$ \Leftrightarrow\qquad E_{\pm\frac 12} \cong \sig\vert_{x=0}P_\pm + 
\mu\vert_{x=0}P_\mp , \eqno(1.2b) $$
where $P_\pm$ are the projections onto the subspaces of integer and 
half-integer charge. (For the precise notations see Sect.\ 2.1 below.)

The general theory says that the restriction of a Wightman field in $d$ 
dimensions to a time-like hyperplane (such as $x^1=0$) yields another 
Wightman field in one dimension less \cite\B. For $d=1+1$, the resulting 
field is a chiral field; the ``positive'' direction (pertinent to the 
spectrum condition) is inherited from the forward time-like direction. 

In order to prevent confusion, we note that the commutation relations obtained 
by restricting a 1+1-dimensional field to the line $x=0$ of course test the 
time-like commutation relations of that field. There is no conflict between 
$\sig$ and $\mu$ being local fields and their representations (1.2a) in 
terms of non-local vertex operators.

The second input is the well-known fact that the (local, but not mutually 
local) 1+1-dimensional order and disorder fields $\sig$ and $\mu$ possess a 
bilinear factorization into chiral exchange fields $a$ and $b$ along with 
their adjoints \cite\RS. The latter, often also called non-abelian vertex 
operators referring to the non-abelian superselection structure of the 
underlying chiral observables (here: the stress-energy tensor $T$ with central 
charge $c = \frac 12$), are the primary fields of scaling dimension 
$h=\frac 1{16}$ for the latter \cite\BPZ, making different transitions 
between the three sectors of $T$. We shall call $a$, $b$, $a^+$ and $b^+$ 
the ``elementary fields'' in the sequel. They are non-local Wightman fields.
(Still, they are relatively local with respect to the chiral observables;
in this sense, all non-local fields throughout this article are not as 
non-local as could be. Genuinely non-local fields without any organizing 
local observables might be far more intricate objects.)

We have to distinguish the factorization in two independent light-cone 
coordinates $u = t+x$, $v = t-x$, as mentioned here, from $p$-factorization
described before. Let $\Phi$ and $\Psi$ be two chiral (non-local) Wightman 
fields; then the $\times$-product
$$ (\Phi \times \Psi)(f) = \int dt\,dx f(t,x) \; \Phi(t+x) \otimes \Psi(t-x) 
\eqno(1.3) $$
defines a 1+1-dimensional (a priori also non-local) Wightman field. The 
$\times$-product is just the ordinary tensor product of operator valued 
distributions (except for the passage to light-cone coordinates), and is 
therefore the tensor product both as a distribution and as an operator.

The bilinear factorization of the order and disorder fields is given in
eqs.\ (3.1) below. For the moment, the gross structure 
$$ (\sig \text{and} \mu) \sim (\hbox{linear}) \times (\hbox{linear}) 
\eqno(1.4) $$
is sufficient. (Here and in the sequel expressions like ``linear'' or 
``quadratic'' mean ``linear or $p$-quadratic polynomial in the elementary 
fields''.)

The point is now that the $\times$-product, when restricted along a time-like 
axis (such as $x=0$), turns into the $p$-product; symbolically:
$$ (\Phi \times \Psi)(t,0) = (\Phi \odot \Psi)(t) . \eqno(1.5) $$
Thus, restricting a 1+1-dimensional field which is obtained as a 
$\times$-product of chiral fields produces the $p$-product of the chiral 
fields.

We apply this observation to the decomposition (1.4) and insert it into the
Kadanoff-Ceva result (1.2b). This yields a representation of the chiral 
abelian vertex operators of charge $\pm\frac 12$ (scaling dimension 
$h=\frac 18$) as a quadratic $p$-polynomial in the (non-local and non-abelian)
elementary fields of scaling dimension $h=\frac 1{16}$
$$ E_{\pm\frac 12} \sim (\hbox{quadratic in the elementary fields}) . 
\eqno(1.6) $$

The exact formula, eq.\ (3.4), is the basic result of the present note. While 
it is quite surprising as it stands, it becomes even more interesting by its 
consequences. Namely, we apply it in several ways by exhibiting various 
familiar fields of various chiral and two-dimensional models as homogeneous 
$p$- and $\times$-product combinations in the elementary fields $a$ and $b$; 
we thereby reproduce several well-known observations from a unifying point of 
view, as well as produce several new results.

1. In Sect.\ 4.1 we give another proof for the massless case of the 
previously mentioned observation by Schroer and Truong \cite\ST\ in the 
``doubled'' Ising model. It relates the $p$-squares of the order and disorder 
fields to the axial potential of the massless Dirac theory. The latter
has the form $\varphi_R(t-x) - \varphi_L(t+x)$, so its exponentials are 
$\times$-product of (oppositely charged) chiral abelian vertex operators. The 
identifications given in \cite\ST\ are the sine and cosine formulae
$$ \sig_D \equiv \sig \odot \sig \cong \frac1{\sqrt2i} 
(E_{-\frac 12} \times E_{\frac 12}  - E_{\frac 12} \times E_{-\frac 12}) , 
\eqno(1.7a) $$
$$ \mu_D \equiv \mu \odot \mu \cong \frac1{\sqrt2} (E_{-\frac 12} 
\times E_{\frac 12} + E_{\frac 12} \times E_{-\frac 12}) . \eqno(1.7b) $$
Since $\sig$ and $\mu$ are $\times$-bilinear in the elementary fields, their 
$p$-squares are $\times$-products of $p$-quadratic expressions in the 
elementary fields. The latter arrange in such a way that the basic result 
turns them into abelian chiral vertex operators, thus reproducing eqs.\ (1.7)
(up to a unitary similarity transformation, and only 
on the joint cyclic Hilbert space of the fields under consideration).

2. We apply our methods to conformal coset models \cite\GKO. 
In these models, the primary fields with respect to a current algebra are 
factorized into primary fields with respect to some current subalgebra and 
primary fields with respect to the corresponding coset algebra 
(``branching''). This separation of degrees of freedom is another example of 
$p$-factorization.

In particular, we consider the U(2) currents of a doublet of complex chiral 
fermions (the primary fields), and its subalgebra of SU(2) currents at level 
1 (Sect.\ 2.3.). The separation of the SU(2) and U(1) (= coset)
degrees of freedom of the complex fermion doublet \cite\BRSS\ is an instance 
of the trivial abelian $p$-factorization of abelian vertex operators as 
mentioned before. It gives rise to the pseudo-real doublet of SU(2)-primary 
fields represented in terms of chiral abelian vertex operators with scaling 
dimension $h=\frac 14$.

In the next step (Sect.\ 4.2.), the latter are $p$-factorized further into
two abelian vertex operators with scaling dimension $h=\frac 18$, 
which in turn are $p$-quadratic in the elementary fields by the basic result. 
The resulting $p$-quartic expression for the level 1 primary fields, when 
written in the form (linear) $\odot$ (cubic), is precisely the coset model 
factorization with respect to the coset stress-energy tensor (with central 
charge $c=\frac 12$) and the level 2 currents, when the level 2 theory is 
embedded into the tensor product of two level 1 theories \cite\GKO. 

The primary chiral exchange fields of scaling dimension $h=\frac 3{16}$ at 
level 2 are pseudo-real isospin doublets of non-abelian exchange fields 
$A_i$ and $B_i$ along with their adjoints, making transitions between the 
three sectors of the SU(2) current algebra. The previous discussion yields a 
$p$-cubic representation in terms of the elementary fields 
$$ A_i \text{and} B_i \sim (\hbox{cubic in the elementary fields}) . 
\eqno(1.8) $$ 
The exact formulae are eqs.\ (4.12) below. This representation explains the 
observation that the solutions to the Knizhnik-Zamo\-lodchikov equation 
\cite\KZ\ for isospin $\frac 12$ at level 2 are cubic expressions in the 
$h=\frac1{16}$ conformal blocks of the $c = \frac 12$ minimal model \cite\BPZ.

3. In Sect.\ 4.3 we consider the defining matrix field $g_{ij}$ of the 
Wess-Zumino-Witten (WZW) model (= generalized Thirring model) \cite\WZW\ with 
$\SU_L \times \SU_R$ symmetry. It is $\times$-bilinear in the primary chiral 
exchange field doublet ($A_i$ and $B_i$ and their adjoints) of the level 2 
SU(2) current algebra, just as $\sig$ and $\mu$ factorize into the chiral 
$a$ and $b$ fields.

According to our previous result (1.8), the chiral exchange field doublets are 
$p$-cubic in the elementary fields, and hence $g_{ij}$ are of the form 
(cubic) $\times$ (cubic). It is possible to split off a $p$-quadratic part in 
the form of abelian vertex operators $E_{\pm\frac12}$ in either chiral factor, 
and the remaining $p$-factor (linear) $\times$ (linear) can be identified 
with $\sig$ and $\mu$ for the diagonal and off-diagonal matrix elements.
The gross structure is
$$ g_{\hbox{\trm diag}} \sim \sig \odot (E_{\pm\frac12} \times E_{\mp\frac12})
, \eqno(1.9a) $$ 
$$ g_{\hbox{\trm offd}} \sim \mu \odot (E_{\pm\frac12} \times E_{\pm\frac12}) 
. \eqno(1.9b) $$ 
The trace of $g$ involves $\sig$ as well as a field which we already know 
how to identify with $\sig \odot \sig$ by eq.\ (1.7a); hence
$$ \hbox{Tr}\, g \cong \sqrt2 \; (\sig \odot \sig \odot \sig) . 
\eqno(1.10) $$

Next, we restrict $g_{ij}$ to a time-like axis and obtain a $p$-polynomial 
of order six in the elementary fields. By eqs.\ (1.2a) and (1.9), these
polynomials turn into cubic $p$-products of vertex operators 
$E_{\pm\frac 12}$. The resulting representation (4.19) of $g_{ij}\vert_{x=0}$ 
by abelian chiral vertex operators with scaling dimension $\frac 38$
generalizes Kadanoff's result (1.2a) to the WZW model.

4. In Sect.\ 4.4 we consider $p$-quadratic expressions in the 1+1-dimensional 
matrix field $g_{ij}$, notably the $p$-determinant. In terms of the 
elementary fields, this determinant field is of the form (order six) 
$\times$ (order six). Again arranging $p$-quadratic polynomials in either 
chiral factor to yield abelian vertex operators according to the basic 
result, and performing some astute similarity transformations, the 
$p$-determinant field is finally found to be 
$$ \hbox{Det}_\odot\, g \cong \frac1{\sqrt2i}(E_{-\b} \times E_\b - 
E_\b \times E_{-\b}) \eqno(1.11) $$ 
with charge $\b = \frac12 \sqrt3$ (rather than $\frac12$ in eq.\ (1.7a)). 
This generalizes the result of Schroer and Truong \cite\ST\ (in the massless
case) to the WZW model. 

A technical comment might be in order. Throughout the paper, we prove
identities between various fields only ``on the cyclic Hilbert space, and up 
to a unitary similarity transformation which preserves the vacuum vector''. 
This means, of course, no weakening of the statement, since the vacuum 
expectation values which determine the fields are never affected. A system 
of fields must of course be subjected to the same transformation.
One should, however, keep in mind other fields which are possibly present, 
and which might or might not be transformed at the same time. Notably, the 
property of being primary with respect to some chiral stress-energy tensor or 
current algebra is defined by specific commutation relations with these chiral 
observables, and will be preserved only if the latter are. This is, e.g., 
guaranteed if the unitary operator which implements the similarity 
transformation is a constant on each irreducible representation space of the 
chiral observables. Otherwise, the latter would have to be transformed
along with the primary fields. Even if the state of affairs is not explicitly 
mentioned in every single instance, it is contextually evident everywhere 
in the paper
by inspection of the specific form of the similarity transformation at hand.

The present work contains the results of the diploma thesis of the first 
author (K.F.) \cite\KF.

\vskip5mm {\bbf 2. Preparation of the ground} \vskip3mm

{\bf 2.1. Abelian vertex operators} \vskip2mm

Unlike the massless free scalar field in 1+1 dimensions, its exponentials 
exist on a positive definite Hilbert space \cite{\W,\LS,\SS,\CRW}. Mandelstam 
\cite\M\ has introduced such exponentials depending on only one chiral 
coordinate when he bosonized the massless Thirring model (at vanishing 
coupling constant) in terms of the sine-Gordon model (at $\b^2 = 4\pi$). 
These fields, denoted by $E_\a = \triple{\exp{i\a\varphi}}$ in the sequel,
are now known as (abelian) chiral vertex operators. The ``triple ordering'' 
refers to the creation and annihilation parts of the auxiliary chiral 
scalar field and involves a renormalization parameter $\mu$ of the dimension 
of mass which is understood in the limit $\mu \searrow 0$. 
For rather streamlined treatments see, e.g., \cite\SS\ and \cite\BB.

The complex free chiral fermion field $\psi$ can be identified with a vertex 
operator of unit charge $\a=1$. Its current $j = \double{\psi\psi^*} = 
\partial\varphi$ generates a U(1) symmetry with charge operator 
$Q = \frac1{2\pi}\int j(x)dx$. By virtue of
$$ QE_\a = E_\a(Q+\a) , \eqno(2.1) $$
the U(1) symmetry extends to the chiral vertex operators of arbitrary 
charge $\a$.

Chiral vertex operators of general charge are non-local chiral Wightman 
fields, satisfying anyonic commutation relations
$$ E_\a(x) E_\b(y) = \eu^{\pm i \pi\a\b} E_\b(y) E_\a(x) 
\text{if} x \neq y \eqno(2.2) $$
with the sign $\pm = \hbox{sign}\;(x-y)$ in the exponent. They are 
conformally covariant fields with scaling dimension $h = \frac12\a^2$ with 
respect to the stress-energy tensor $\frac i{4\pi}\double{\psi\ppartial\psi^*} 
= \frac1{4\pi}\double{j^2}$ of the complex fermion (with central charge $c=1$).

The vacuum expectation values of chiral vertex operators satisfy
charge conservation and are given by
$$ \vac{E_{\a_1}(x_1) \ldots E_{\a_n}(x_n)} = \prod_{i<j} 
\Delta(x_i-x_j)^{-\a_i\a_j} \quad\text{provided} \sum \a_i = 0\;. \eqno(2.3) $$
The distributions 
$$ \Delta(x-y)^{2h} = \left[\frac{-i}{x-y-i\epsilon}\right]^{2h} = 
\frac 1{\Gamma(2h)} \int_0^\infty dk k^{2h-1} \eu^{-ik(x-y)} \eqno(2.4) $$
contributing to these v.e.v.'s have spectral support on the positive momenta.

Abelian chiral vertex operators with multi-component charges can be defined as 
exponentials of several independent massless free fields, e.g.,
$$ E_{(\a,\b)} = \triple{\exp i(\a\varphi_1 + \b\varphi_2)} \cong 
E_\a \odot E_\b \; . \eqno(2.5) $$
Their vacuum expectation values are given by the same formula (2.3), 
replacing only products of charges by scalar products of multi-component 
charges. This replacement results in the factorization of the total v.e.v.'s 
into the v.e.v.'s for the charge components and therefore justifies the  
equivalence in eq.\ (2.5). 

On the other hand, it implies an obvious O($N$) symmetry for the vertex 
operators with $N$-component charges, corresponding to rotations and
reflections in charge space. Therefore on its own cyclic Hilbert space, the 
field (2.5) is also equivalent to $E_{\sqrt{\a^2+\b^2}}$.

\vskip5mm {\bf 2.2. Exchange fields} \vskip2mm

Chiral exchange fields arise by a spectral decomposition with respect to 
global conformal transformations of 1+1-dimensional fields with conformal 
symmetry \cite\SSV\ and subsequent chiral factorization. They arise also  
as point-like limits of bounded intertwining operators in the 
framework of algebraic quantum field theory \cite{\FRS,\J}, or as intertwiners 
between modules of chiral symmetry algebras such as Virasoro or affine 
Kac-Moody algebras \cite\VO. They are the fundamental charge-carrying 
entities which make transitions between the various superselection sectors of 
a theory of local chiral observables. As such, they are defined on a 
``source'' Hilbert space and take values in a ``range'' Hilbert space which
are both irreducible sectors of the observables, and are relatively local with
respect to the latter.
 
In specific models where the chiral observables are generated by the 
stress-energy tensor or by a current algebra, it is convenient to distinguish
``primary'' exchange fields by their specific commutation relations with the
observables. These imply certain partial differential equations, reflecting 
the symmetry of the model, for the vacuum expectation values of primary 
exchange fields whose solutions are known as ``conformal blocks'' 
\cite{\BPZ,\KZ}.
The commutation relations of chiral exchange fields with each other are given 
by a non-abelian representation of the braid group. 
Abelian vertex operators (restricted to a charge sector) are the exchange 
fields with respect to an abelian current algebra.

The primary chiral exchange fields of the Ising model were first treated in 
detail in \cite\RS. In this model, the chiral observables are generated by
the stress-energy tensor with central charge $c=\frac 12$. There are three 
superselection sectors, which carry representations of the conformal group with
primary scaling dimensions $0$, $\frac 1{16}$ and $\frac 12$, respectively.

The primary field for the trivial sector $0$ is just the unit operator.
The primary field for the sector $\frac12$ is a real free fermion. There are
three exchange fields corresponding to this primary field, which are
given by the free fermion field in the Neveu-Schwarz representation 
restricted to the subspaces of even and odd Fermi number, and the free 
Fermi field in the Ramond representation. The former two are each other's
adjoints $\psi_{\hbox{\trm NS}}: \hh_0 \arr \hh_{\frac12}$ and 
$\psi_{\hbox{\trm NS}}^+: \hh_{\frac12} \arr \hh_0$, and the latter is a 
selfadjoint field $\psi_{\hbox{\trm R}}: \hh_{\frac1{16}} \arr 
\hh_{\frac1{16}}$.

The primary exchange fields for the sector $\frac1{16}$ are the chiral 
constituents of the order and disorder parameter of the 1+1-dimensional
Ising model. There are two ``raising'' operators $a$ and $b$ along with their 
adjoints:
$$ a: \hh_0 \arr \hh_{\frac1{16}} \text{and its adjoint} a^+: \hh_{\frac1{16}}
 \arr \hh_0  \eqno(2.6a) $$
$$ b: \hh_{\frac1{16}} \arr \hh_{\frac12} \text{and its adjoint} 
b^+: \hh_{\frac12} \arr \hh_{\frac1{16}} \eqno(2.6b) $$

The commutation relations of these chiral exchange fields are given by a
non-abelian representation of the braid group of the Hecke type \cite\RS.

We have chosen the normalization such that the two-point function is
$\vac{a^+(x)a(y)} = \Delta(x-y)^{\frac18}$. The only possible four-point 
functions are given by 
$$ \vac{a^+(x_1)a(x_2)a^+(x_3)a(x_4)} = V^{\frac 18} \cdot f(x) \eqno(2.7a) $$
$$ \vac{a^+(x_1)b^+(x_2)b(x_3)a(x_4)} = V^{\frac 18} \cdot g(x) \eqno(2.7b) $$
where the first factor with 
$$ V(x_1,x_2,x_3,x_4) = 
\frac{\Delta(x_1-x_2)\Delta(x_3-x_4)\Delta(x_1-x_4)\Delta(x_2-x_3)}
{\Delta(x_1-x_3)\Delta(x_2-x_4)} \eqno(2.8) $$
is the four-point function of abelian vertex operators with alternating 
charges and the proper scaling dimension, and the ``conformal block'' functions
$$ f(x) = \sqrt{\textstyle \frac12(1+\sqrt{1-x})} \text{and} 
g(x) = \sqrt{\textstyle \frac12(1-\sqrt{1-x})} \eqno(2.9) $$
depend only on the conformally invariant cross ratio  
$x = \frac{\Delta(x_1-x_3)\Delta(x_2-x_4)}{\Delta(x_1-x_2)\Delta(x_3-x_4)}$.

The higher $2n$-point functions were derived in closed form in \cite\RS.

\vskip5mm {\bf 2.3. The SU(2) primary fields at level 1 in terms of vertex 
operators} \vskip2mm

A doublet of complex free chiral fermions $\psi_\pm$ gives rise to a theory
with U(2) symmetry. Its stress-energy tensor with $c=2$ is given by
$T = \frac i{4\pi}(\double{\psi_+\ppartial\psi_+^*} + 
\double{\psi_-\ppartial\psi_-^*})$, and the currents are
$$ j^A = \double{\psi_i\tau^A_{ij}\psi_j^*} \qquad (A = 0,1,2,3) 
\eqno(2.10) $$
where $\tau^0 = \eins$ and $\tau^a$ ($a=1,2,3$) are the Pauli matrices.

The component $j^0$ generates the abelian U(1) current algebra with 
stress-energy tensor
$T_U = \frac1{8\pi}\double {(j^0)^2}$ ($c=1$), and $j^a$ generate the SU(2) 
current algebra at level 1 with stress-energy tensor $T_S$ given by the 
Sugawara formula quadratic in the currents ($c=1$), such that $T = T_U + T_S$. 
The U(1) and the SU(2) currents commute with each other, and both transform 
the fermion doublet in the obvious way.

The U(1) and the SU(2) degrees of freedom of the fermion doublet can be 
separated as follows in a very explicit manner \cite\BRSS. The crucial point 
is the fact that chiral charged fermions can be represented by abelian vertex 
operators of unit charge.

The doublet of charged chiral fermions can be represented by a 
pair of abelian vertex operators with orthogonal unit charge vectors
$\a_\pm$. Since the latter constitute a pair of commuting fermions, a Klein 
transformation is needed to make them anti-commute. Rather than the standard 
choice (multiplying, e.g., $E_{\a_+}$ by the exponential $\exp(i\pi\a_-Q)$ of 
the charge operator of the other component), we prefer a more symmetric one 
and multiply both components by $\exp(i\pi\frac{\a_+ - \a_-}2Q)$. Both
choices are unitarily equivalent by conjugation with the exponential of some
quadratic polynomial of the charge operators.

We choose $\a_\pm = \frac1{\sqrt2}\cdot(1,\pm1)$ and
apply the $p$-factorization of multi-component vertex operators
(eq.\ (2.5)). This gives rise to the simple $p$-factorization 
$$ \psi_\pm = \phi \odot \phi_\pm \eqno(2.11) $$
where
$$ \phi = E_\a \eqno(2.12) $$ 
is an ordinary abelian vertex operator of charge $\a = \frac1{\sqrt2}$ 
carrying the U(1) symmetry, and 
$$ \phi_\pm = E_{\pm\a} \; \eu^{i\pi\a Q} \eqno(2.13) $$ 
is a pseudoreal doublet carrying the SU(2) symmetry. Pseudoreality means
$$ \phi_i^* = i\eps_{ij} \; \phi_j C = -i\eps_{ij} \; C \phi_j \eqno(2.14) $$ 
where $\eps_{+-} = 1$ and $C = \eu^{2\pi i\a Q} = C\inv$ is a Casimir operator 
which equals $+1$ ($-1$) on all states of integer (half-integer) isospin and 
thus commutes with the SU(2) symmetry.

In this representation, the factorization of the current algebra becomes
manifest: the U(1) current is $j^0 = \sqrt2 \; \partial \varphi \otimes 
\eins$, and the SU(2) currents are found in the Frenkel-Kac \cite\KF\
representation $j^3 = \eins \otimes \sqrt2 \; \partial \varphi$ and 
$j^\pm = \eins \otimes E_{\pm\sqrt2}$.

Both the U(1) field and the SU(2) doublet are primary fields w.r.t.\ the 
respective current algebra, and have scaling dimension $h=\frac 14$. They
both satisfy anyonic commutation relations according to eq.\ (2.2).
The two-point functions are $\vac{\phi^*\phi} = \Delta^{\frac12}$, or 
equivalently $\vac{\phi_i\phi_j} = -i\eps_{ij} \Delta^{\frac12}$.

The cyclic Hilbert space of the pseudoreal doublet contains all charges 
which are integer multiples of $\a=\frac 1{\sqrt 2}$. The  primary exchange 
fields which interpolate the two sectors of the SU(2) current algebra at 
level 1 (integer and half-integer isospin), are obtained by restricting 
$\phi_\pm$ to the subspaces of even or odd charge in units of $\a$, or
equivalently of integer or half-integer isospin.

\newpage
\vskip5mm {\bf 2.4. Goddard-Kent-Olive (GKO) coset construction} \vskip2mm

Consider two current algebras $\aa_i$ for some simple Lie group $G$ at level 
$k_i$ generated by the currents $j_i^a$, and with stress-energy tensor $T_i$
given by the affine Sugawara formula. The diagonal currents
$$ J^a = j_1^a \otimes \eins + \eins \otimes j_2^a \eqno(2.15) $$
generate a current algebra $\aa$ at level $k_1+k_2$ within $\aa_1 \otimes
\aa_2$. This current algebra has its own stress-energy tensor $T$ given by 
the Sugawara formula in terms of $J^a$ which has the same commutation 
relations with $J^a$ as the total stress-energy tensor $T_1 \otimes \eins + 
\eins \otimes T_2$. The difference $T_{\hbox{\trm coset}}$ is a stress-energy 
tensor which commutes with the current algebra $\aa$. It belongs 
(possibly along with other operators) to the coset model $\cc$ of fields
in $\aa_1 \otimes \aa_2$ commuting with $\aa$.

Embedding a primary exchange field at level $k_1$ by the prescription 
$\phi \otimes \eins$ into the Hilbert space of $\aa_1 \otimes \aa_2$, yields 
operators which have primary commutation relations with both the diagonal 
currents $J^a$ and the coset stress-energy tensor. Therefore, according to 
the same scheme of separation of degrees of freedom as in Sect.\ 2.3, there 
is a factorization of the form
$$ \phi \otimes \eins \cong \sum \varphi_c \odot \varphi \eqno(2.16) $$
where $\varphi_c$ are primary exchange fields w.r.t.\ the coset theory and
$\varphi$ are primary exchange fields w.r.t.\ the current algebra at level
$k_1+k_2$. The quantum numbers of the exchange fields on the right hand side
are determined by the branching rules of representations of
$\aa_1\otimes\aa_2$ in restriction to $\cc\otimes\aa$; in particular,
$\varphi$ carries the $G$ quantum numbers, and the scaling dimensions
are additive.

The symbol $\odot$ on the right hand side of the GKO decomposition (2.16) 
refers to the tensor product of representation spaces of $\cc \otimes \aa$, 
while in contrast, on the left hand side $\otimes$ refers to the tensor 
product $\aa_1 \otimes \aa_2$. The passage is made by a reorganization of the
Hilbert space which, in a special case, will become explicit along the way
in Sect.\ 4.2.

\vskip5mm {\bbf 3. The basic result} \vskip3mm

Chiral exchange fields can be combined in various ways to yield local 
1+1-dimensional fields. A systematic way to do so was described in 
\cite\Spt\ in terms of the algebraic theory of superselection sectors.
(Helas, the assignment of $\sig$ and $\mu$ was interchanged in \cite\Spt,
and no care was taken to construct local fields which are closed under 
conjugation; in order to repair that defect, the self-adjoint local fields 
below differ from the result in \cite\Spt\ by a non-unitary similarity 
transformation. One can actually adjust the correct formulae ``by hand'' from
the known commutation relations and four-point functions of the elementary 
$a$ and $b$ fields \cite\RS\ by imposing the correct commutation relations 
\cite\ST\ and four-point functions \cite\RS\ for $\sig$ and $\mu$.)

The upshot is the following. The local field $\sig$ interpolates between 
the ``diagonal'' charge sectors $\hh_{ss} \equiv \hh_s \otimes \hh_s$ where 
$\hh_s$ ($s=0,\frac1{16},\frac12$) are the three superselection sectors of 
the chiral observables. It is given by the standard formula
$\sig = a \times a + b \times b + (\hbox{h.c.})$. The field $\mu$, however,
carries an ``excess $\ZZ_2$ charge'' which makes transitions to the
``non-diagonal'' sectors $\hh_{0\frac12}$, $\hh_{\frac12 0}$ and another copy 
of $\hh_{\frac1{16}\frac1{16}}$.
In order to describe both fields simultaneously, $\sig$ has of course
to be defined also on the non-diagonal sectors. There is some ambiguity
how to define these fields which consists in unitary similarity transformations
which commute with the chiral observables in order not to spoil the
primary commutation relations. These unitaries are given by a complex
phase on each of the four simple sectors, and a unitary 2$\times$2 matrix 
which mixes the two copies of $\hh_{\frac1{16}\frac1{16}}$.

Our representation of choice is the following.
$$ \sig = u \otimes (a \times a + b^+ \times b^+) + 
v \otimes (a \times b^+ + b^+ \times a) + (\hbox{h.c.}) , \eqno(3.1a) $$ 
$$ \mu = v \otimes (a \times a - b^+ \times b^+) + 
iu \otimes (a \times b^+ - b^+ \times a) + (\hbox{h.c.}) . \eqno(3.1b) $$ 
We introduced two orthogonal unit vectors $u$ and $v$ in $\CC^2$ in order to 
distinguish the two copies of $\hh_{\frac1{16}\frac1{16}}$ within 
$\hh_{\frac1{16}\frac1{16}} \oplus \hh_{\frac1{16}\frac1{16}} = \CC^2 \otimes 
\hh_{\frac1{16}\frac1{16}}$; they are multiplied like rectangular matrices.
Obviously, $\sig$ preserves both $\hh_{00} \oplus (u \otimes 
\hh_{\frac1{16}\frac1{16}}) \oplus \hh_{\frac12\frac12}$ and $\hh_{0\frac12} 
\oplus (v \otimes \hh_{\frac1{16}\frac1{16}}) \oplus \hh_{\frac12 0}$, while 
$\mu$ alternates between these spaces. Conversely, $\mu$ preserves the 
analogous spaces with $u$ and $v$ interchanged, while $\sig$ alternates 
between them. The following diagram summarizes the situation: $\sig$ 
interpolates according to the horizontal
arrows, while $\mu$ interpolates according to the diagonal arrows: 
$$ \matrix(00) & \longleftrightarrow & u \otimes (\frac1{16}\frac1{16}) &
\longleftrightarrow & (\frac12\frac12) \\ & \cross && \cross & \\ 
(0\frac12) & \longleftrightarrow & v \otimes (\frac1{16}\frac1{16}) &
\longleftrightarrow & (\frac12 0) \endmatrix \eqno(3.2) $$
Every single term in (3.1) corresponds to an arrow in (3.2).

Our normalization is such that the two-point functions are
$\vac{\sig\sig} = \vac{\mu\mu} = (\Delta_L\Delta_R)^{\frac18}$ in terms 
of the left- and right-moving light-cone coordinates. All mixed four-point 
functions \cite\RS, e.g., $\vac{\mu\mu\sig\sig} = (V_LV_R)^{\frac18}
(f_Lf_R - g_Lg_R)$, can be read off eqs.\ (2.7) and (3.1).

Each of the fields $\sig$ and $\mu$ satisfies local commutativity at 
space-like distance with itself, but they are not mutually local.
Instead, they satisfy the ``dual'' commutation relations \cite\ST
$$ \sig(x)\mu(y) = \pm \mu(y)\sig(x) \text{if} (x-y)^2<0 \eqno(3.3) $$
with the sign $\pm = \hbox{sign}\; (x^1-y^1)$.

Let us consider now the Kadanoff-Ceva formulae (1.2). Evaluating the fields 
$\sig$ and $\mu$ at $x=0$ means replacing the $\times$-product in (3.1)
by the $p$-product $\odot$. The projections $P_+$ and $P_-$ project onto 
the subspaces generated from the vacuum by an even resp.\ odd number of 
applications of the fields $\sig$ and $\mu$, i.e., $P_+$ projects onto 
$\hh_+ = \bigoplus_{s,t = 0,\frac12}\hh_{st}$ while $P_-$ projects onto
$\hh_- = \CC^2 \otimes \hh_{\frac1{16}\frac1{16}}$.

Thus inserting (3.1) into (1.2b) yields the basic result
$$ \eqalign{E_{+\frac12} & = u \otimes (a \odot a + b^+ \odot b^+) - 
iu^+ \otimes (a^+ \odot b - b \odot a^+) \cr 
& \;\; + v \otimes (a \odot b^+ + b^+ \odot a) + 
v^+ \otimes (a^+ \odot a^+ - b \odot b)  , \cr}
\eqno(3.4) $$
and $E_{-\frac12} = E_{+\frac12}^*$.

Like all formulae below, eq.\ (3.4) can easily be checked in all mixed 
four-point functions of $E_{\pm\frac12}$, using eqs.\ (2.3) and (2.7). The
important point is, however, its validity on the entire joint cyclic
Hilbert space of these fields.

The charge conjugation invariance $E_{+\frac12} \leftrightarrow E_{-\frac12}$ 
is manifest in this representation. It is implemented by the involutive 
unitary operator $\Pi \cdot (P_{00} + i P_{0\frac12} - i P_{\frac12 0} - 
P_{\frac12\frac12} + M \otimes P_{\frac1{16}\frac1{16}})$ involving 
the discrete ``flip'' operator $\Pi: \hh_{st} \arr \hh_{ts}$ which 
interchanges the two tensor factors, and a matrix $M$ in the multiplicity 
space of $\hh_{\frac1{16}\frac1{16}}$ which interchanges $u$ and $v$.

Eq.\ (3.4) shows a $\ZZ_4$ charge structure as follows.
We redisplay the diagram (3.2) with only the arrows pertaining to
$E_{+\frac12}$:
$$ \matrix (00) & \longrightarrow & u \otimes (\frac1{16}\frac1{16}) &
\longleftarrow & (\frac12\frac12) \\ & \westcross && \eastcross & \\
(0\frac12) & \longrightarrow & v \otimes (\frac1{16}\frac1{16}) &
\longleftarrow & (\frac12 0) \endmatrix \eqno(3.5) $$
It becomes apparent that the subspaces $H_i$ given by
$$ \eqalign{H_0 = \hh_{00} \oplus \hh_{\frac12\frac12} \qquad & \qquad
           H_1 = u \otimes \hh_{\frac1{16}\frac1{16}} \cr
           H_2 = \hh_{0\frac12} \oplus \hh_{\frac12 0} \qquad & \qquad
           H_3 = v \otimes \hh_{\frac1{16}\frac1{16}} \cr} \eqno(3.6) $$
are the subspaces of charge $(i \mod 4)\frac12$. The four terms in 
eq.\ (3.4) correspond in turn to $P_{i+1}E_{+\frac12}P_i = E_{+\frac12}P_i$, 
$i=0,1,2,3$.

In terms of these subspaces we have the interpolation diagrams
$$ \sig: \left\{\matrix H_0  \leftrightarrow  H_1 \\
\quad \\ H_3  \leftrightarrow  H_2 \endmatrix \right. \qquad\qquad
\mu: \left\{\matrix H_0 \qquad  H_1 \\ \updownarrow \qquad \updownarrow 
\\ H_3 \qquad H_2 \endmatrix \right. \qquad\qquad
E_{\frac12}: \left\{ \matrix H_0 \rightarrow  H_1 \\ \uparrow \qquad 
\downarrow \\ H_3 \leftarrow H_2 \endmatrix \right. \eqno(3.7) $$

The first two diagrams pertain to the 1+1-dimensional theory. The $\ZZ_4$
charge equals the Kadanoff-Ceva $\Gamma$-charge which is conserved
``off the line'' only mod 4. In this picture, the real chiral Ising fermions 
$\psi_L$ and $\psi_R$ arising in the operator product expansions of 
$\mu\sig$ and $\sig\mu$ carry two units of $\ZZ_4$ charge.

\vskip5mm {\bbf 4. Applications} \vskip3mm

{\bf 4.1. The doubled Ising model} \vskip2mm

In contrast to the order and disorder fields $\varphi = \sig,\mu$ with 
their dual commutation relations (3.3), the ``doubled'' fields $\varphi_D = 
\varphi \odot \varphi$ satisfy local commutativity also with each other.
The asserted identifications (1.7) which we are going to establish 
are equivalent to 
$$ \mu_D + i \sig_D \cong \sqrt2 \cdot E_{-\frac12} \times E_{\frac12} 
\eqno(4.1) $$
where the fields on the right hand side are also local fields made 
from anyonic chiral constituents \cite\BB. 

We compute $E_{-\frac12} \times E_{\frac12}$ from the basic result (3.4). 
Its cyclic Hilbert space involves only the conjugate charge sectors $H_{-i} 
\otimes H_i$. For the vacuum expectation values, we may therefore omit all 
terms which are defined on $H_j \otimes H_i$, $j \neq -i$. The remaining 
terms are
$$ \eqalign{E_{-\frac12} \times E_{\frac12} \cong 
 Y & \otimes (aa - b^+b^+) \times (aa + b^+b^+) \cr
 + iX & \otimes (ab^+ - b^+a) \times (ab^+ + b^+a) \cr 
 + X^+ & \otimes (a^+a^+ + bb) \times (a^+a^+ - bb) \cr
 - iY^+ & \otimes (a^+b + ba^+) \times (a^+b -ba^+) \cr} \eqno(4.2) $$
on the cyclic subspace. We omitted the $\odot$ symbols, and put 
$X = u \otimes v$, $Y = v \otimes u$.

On the other hand, we compute the doubled fields from eqs.\ (3.1). Each of the 
fields $\sig$ and $\mu$, according to the scheme (3.7), takes each of the 
subspaces
$H_i$ into a specific $H_j$. Hence repeated application of the doubled fields
connects the vacuum sector $H_0 \otimes H_0$ only with the diagonal
sectors $H_i \otimes H_i$, and the latter again exhaust the joint cyclic 
Hilbert space. Thus we obtain 
$$ \eqalign{\mu_D + i \sig_D \cong 
(V+iU) & \otimes (aa \times aa + b^+b^+ \times b^+b^+ +iaa \times b^+b^+ 
+ib^+b^+ \times aa) \cr -
(V-iU) & \otimes (ab^+ \times ab^+ + b^+a \times b^+a -iab^+ \times b^+a
-ib^+a \times ab^+) \cr +
(V^++iU^+) & \otimes (a^+a^+ \times a^+a^+ + bb \times bb +ia^+a^+ \times bb 
+ibb \times a^+a^+) \cr -
(V^+-iU^+) & \otimes (a^+b \times a^+b + ba^+ \times ba^+ -ia^+b \times ba^+ 
-iba^+ \times a^+b) \cr} \eqno(4.3) $$
on the joint cyclic subspace. We put $U = u \otimes u$, $V = v \otimes v$
and arranged the factors as $(A \times B) \odot (C \times D) \cong 
(A \odot C) \times (B \odot D)$. Note that due to the different factor 
ordering, the individual subspaces $H_j \otimes H_i$ in eq.\ (4.2) 
are different from $H_j \otimes H_i$ in eq.\ (4.3).

Eq.\ (4.3) is transformed into $\sqrt2$ times eq.\ (4.2) by conjugation with 
a unitary operator which preserves the vacuum vector. The operator which does 
the job equals  (in an obvious notation) 
$(P_{00\times 00} - P_{\frac12\frac12\times\frac12\frac12}
+ P_{0\frac12\times\frac12 0} - P_{\frac12 0\times 0\frac12}) 
+i (P_{00\times\frac12\frac12} - P_{\frac12\frac12\times 00}
+ P_{0\frac12 \times 0\frac12} - P_{\frac12 0\times\frac12 0})
+ M \otimes P_{\frac1{16}\frac1{16}\times\frac1{16}\frac1{16}}$, with a
unitary matrix $M$ in the multiplicity space of $\hh_{\frac1{16}\frac1{16}}
\otimes \hh_{\frac1{16}\frac1{16}}$ taking $V+iU \mapsto \sqrt2\, Y$ and 
$V-iU \mapsto \sqrt2\, X$. This similarity transformation commutes with the 
chiral observables of the doubled Ising field theory (two stress-energy 
tensors with central charge $c=\frac12$ on each light-cone), and therefore 
preserves the structure of the full model.

It might be objected that one obtains apparently different results when one 
restricts the doubled fields according to eqs.\ (1.7) to the line $x=0$, and 
when one doubles the restricted fields according to eqs.\ (1.2a); namely
$$ \eqalign{\textstyle \sig_D\vert_{x=0} = \frac1{\sqrt2i}(E_\a - E_{-\a}) \cr 
\textstyle \mu_D\vert_{x=0} = \frac1{\sqrt2}(E_\a + E_{-\a})\cr}
\quad \text{vs.} \quad 
\eqalign{ (\sig\vert_{x=0})_D = E_\a P_+ + E_{-\a} P_- \cr 
(\mu\vert_{x=0})_D = E_{-\a} P_+ + E_\a P_- \vrule width0pt height1.3em \cr} 
\eqno(4.4) $$
where $P_\pm$ now project on the spaces of even resp.\ odd charge in units 
of $\a = \frac1{\sqrt2}$. In fact, the latter form can be transformed 
into the former by a chain of similarity transformations exploiting SU(2) 
invariance. It is implemented by the unitary operator ${\Cal V} 
\eu^{-i\pi(\a Q)^2} {\Cal U} \eu^{i\pi(\a Q)^2}$ where first 
$\eu^{i\pi(\a Q)^2}$ transforms the abelian vertex operators into the 
pseudoreal SU(2) doublet $\phi_\pm$ along with the SU(2)-invariant Casimir 
operator $C = P_+ - P_-$ (cf.\ Sect.\ 2.3), next the SU(2) transformation 
${\Cal U}$ takes $\phi_\pm \mapsto \sqrt{\frac{\mp i}2}(\phi_+ \mp \phi_-)$, 
then $\eu^{-i\pi(\a Q)^2}$ reproduces the vertex operators, and finally 
${\Cal V} = P_+ + \sqrt{-i} P_-$ (which commutes with the other three 
operators) removes the charge dependence. We refrain 
from an explicit demonstration, since a similar transformation will be 
discussed in detail later (Sect.\ 4.4). 

\vskip5mm {\bf 4.2. The SU(2) primary fields at level 2} \vskip2mm

We consider the GKO coset $(\SU_1 \times \SU_1)/\SU_2$.
The coset algebra is the chiral stress-energy tensor with $c=\frac12$
whose primary exchange fields were described in Sect.\ 2.2. 

The level 2 current algebra has also three sectors $\kk_I$ distinguished by 
the isospin $I=0,\frac12,1$ of the ground states of the conformal Hamiltonian.
The primary exchange fields of isospin $I=\frac12$ are four SU(2) doublets 
which make transitions in analogy to eqs.\ (2.6):
$$ A_i: \kk_0 \arr \kk_{\frac12} \text{and} A_i^+: \kk_{\frac12} \arr \kk_0 
\eqno(4.5a) $$
$$ B_i: \kk_{\frac12} \arr \kk_1 \text{and} B_i^+: \kk_1 \arr \kk_{\frac12} 
\eqno(4.5b) $$
The fields $A_i^+$ and $B_i^+$ are not the adjoints of $A_i$ and $B_i$, since
the latter transform in the conjugate representation. We may, however,
choose the normalizations such that 
$$ X_i^* = -\eps_{ij}X_j^+, \qquad (X_i^+)^* = \eps_{ij}X_j \qquad (X = A,B,  
\quad i,j = \pm). \eqno(4.6) $$ 
Like the elementary fields, the chiral exchange fields of the SU(2)
level 2 current algebra satisfy braid group commutation relations given by
another non-abelian Hecke type representation.

The two-point function is $\vac{A_i^+A_j} = \eps_{ij} \Delta^{\frac38}$.
The only non-vanishing four-point functions of these exchange fields
have the structure $\vac{A_i^+A_jA_k^+A_l} = V^{\frac38} \cdot F_{ijkl}(x)$ 
and $\vac{A_i^+B_j^+B_kA_l} = V^{\frac38} \cdot G_{ijkl}(x)$, with $V$ and 
$x$ as in Sect.\ 2.2. The SU(2) tensors $F$ and $G$ are the two independent
``conformal block'' solutions to the Knizhnik-Zamolodchikov differential 
equation 
$$ \left[ (k+2)\partial_x - \frac{C_{12}}x + \frac{C_{23}}{1-x}\right]W(x) = 0
\eqno(4.7) $$
where $k=2$ is the level and $C_{mn} = (\vec I_m + \vec I_n)^2$ are the 
quadratic SU(2) Casimir operators on the product representations due to the 
$m$-th and $n$-th field entry. The solutions 
$$ \eqalign{ F_{ijkl} &= \;\;\, \eps_{ij}\eps_{kl} \cdot f(f^2+g^2) \;-\;
\eps_{ik}\eps_{jl} \cdot 2fg^2 \cr G_{ijkl} &= -\eps_{ij}\eps_{kl} \cdot 
g(f^2+g^2) \; + \; \eps_{ik}\eps_{jl} \cdot 2f^2g \cr } \eqno(4.8) $$
involve the same functions $f(x)$ and $g(x)$ as in eq.\ (2.9).

Our intention is to give an explanation for the cubic structure of these 
vacuum expectation values in terms of the v.e.v.'s of the elementary fields. 
For this purpose, we recall how the level 2 primary exchange fields can be 
obtained by the GKO coset construction from the level 1 primaries 
$\phi_\pm\otimes\eins$.

According to the general discussion in Sect.\ 2.4., the branching rules 
(in particular, the SU(2) quantum numbers and scaling dimensions) determine 
the possible terms in the coset factorization of $\phi_\pm\otimes\eins$. The 
result is
$$ \phi_i \otimes \eins = a \odot A_i -i a^+ \odot A_i^+
+ b \odot B_i +i b^+ \odot B_i^+ . \eqno(4.9) $$
Here the absolute coefficients 1 in front of the first and third term are 
chosen by a normalization freedom, and the relative coefficients $\pm i$ in 
front of the adjoint terms are determined by the pseudoreality property 
(2.14) of $\phi_i$ along with eq.\ (4.6). Eq.\ (4.9) can be cross-checked
by inserting the four-point functions (2.7) and (4.8) and comparing with
the four-point functions of $\phi_\pm$ which are computed with the help
of eqs.\ (2.13) and (2.3). 

Now comes the crucial step. According to eq.\ (2.13), $\phi_\pm$ have a 
representation in terms of vertex operators of charge $\pm\a$ with $\a^2 =
\frac12$. According to eq.\ (2.5) with the choice $\a = (\frac12,\frac12)$, 
the latter can be written in the form $E_{\pm\frac12} \odot E_{\pm\frac12}$, 
and according to eq.\ (3.4), the vertex operators of charge $\pm\frac12$ can 
be written as $p$-quadratic polynomials in the elementary exchange fields. 
The operator $\eu^{i\pi\a Q}$ in (2.13) takes values $i^p$ in the sectors 
of charge $p\a$, and the charge quantum numbers $p \mod 4$ in 
the representation (3.4) have been described in eqs.\ (3.5) and (3.6). Since 
in the cyclic Hilbert space of $E_{\pm\frac12} \odot E_{\pm\frac12}$ only 
these ``diagonal'' charges occur, the mixed terms can be omitted from the 
$p$-square of (3.4). The result is
$$ \eqalign{\phi_+ \otimes \eins  \cong  
[u & \otimes (a \odot a + b^+ \odot b^+)]^{\odot 2} 
+ i [-iu^+ \otimes (a^+ \odot b - b \odot a^+)]^{\odot 2} \cr 
- [v & \otimes (a \odot b^+ + b^+ \odot a)]^{\odot 2} 
-i [v^+ \otimes (a^+ \odot a^+ - b \odot b)]^{\odot 2} . \cr} \eqno(4.10) $$
We reorder this expression by separating the first $p$-factor
(and at the same time make it more lucid by omitting the $\otimes$ and
$\odot$ signs):
$$ \eqalign{\phi_+ \otimes \eins \cong 
\quad a\;\; \odot\; & [Ua(aa+b^+b^+) - Vb^+(ab^++b^+a)] \cr  + 
a^+ \odot\; & [-iU^+b(a^+b-ba^+) -iV^+a^+(a^+a^+-bb)] \cr  + 
b\;\; \odot\; & [iU^+a^+(a^+b-ba^+) +iV^+b(a^+a^+-bb)] \cr  + 
b^+ \odot\; & [Ub^+(aa+b^+b^+) - Va(ab^++b^+a)] \cr} \eqno(4.11) $$
where $U = u \otimes u$, $V = v \otimes v$ are just another pair of 
orthogonal unit vectors pointing to two of the four copies of
$\hh_{\frac1{16}\frac1{16}}^{\otimes 2}$ in the representation space of
$E_{\pm\frac12} \odot E_{\pm\frac12}$.

Of course, also the embedded field $1 \otimes \phi_\pm$ has an equivalent 
representation in terms of exchange fields of the coset and level 2 chiral 
observables. It does, however, not live on the same cyclic Hilbert space 
$(\hh_0\otimes\kk_0) \oplus (\hh_{\frac1{16}}\otimes\kk_{\frac12}) \oplus 
(\hh_{\frac12}\otimes\kk_1)$ as $\phi_\pm \otimes 1$. E.g., the field 
$\phi_+\odot\phi_+$ belongs to a triplet of the diagonal SU(2) and has 
scaling dimension $\frac12$. These quantum numbers are found in the
sectors $\hh_0\otimes\kk_1$ and $\hh_{\frac12}\otimes\kk_0$. In fact, the 
joint cyclic Hilbert space contains apart from the mentioned five sectors also
a second copy of $\hh_{\frac1{16}}\otimes\kk_{\frac12}$, where the two
embedded doublets have similar interpolation properties as $\sig$ and $\mu$
in (3.2). We refrain from computing these formulae.

By comparing (4.11) with (4.9), we read off the $p$-cubic representations
of the level 2 primary exchange doublet:
$$ \eqalign{A_+ &= Ua(aa+b^+b^+) - Vb^+(ab^++b^+a) \cr
A_+^+  &= U^+b(a^+b-ba^+) + V^+a^+(a^+a^+-bb) \cr
B_+  &= iU^+a^+(a^+b-ba^+) +iV^+b(a^+a^+-bb) \cr
B_+^+  &=  -iUb^+(aa+b^+b^+) +iVa(ab^++b^+a) \cr } \eqno(4.12) $$
while the $-$ components are determined by (4.6).

This interpretation of (4.11) means that we choose the embedding in such a 
way that the coset stress-energy tensor (with $c=\frac12$) acts on the 
Hilbert spaces belonging to the first $p$-factor. Then the Sugawara 
stress-energy tensor of the level 2 current algebra (with $c=\frac32$) acts 
on the triple tensor products of Hilbert spaces belonging to the remaining 
three $p$-factors in the form of three commuting $c=\frac12$ 
``stress-energy tensors''. The latter, of course, do not individually 
generate the conformal transformations of any level 2 fields, nor are the 
level 2 currents defined on the individual factor Hilbert spaces; the true
stress-energy tensor with $c=\frac32$ is the sum of the three $c=\frac12$ 
fields.

A non-trivial cross-check of our result (4.12) is that it reproduces the 
four-point functions (4.8) (using eq.\ (2.7)).

Finally, for later use, we display another representation which amounts to
viewing the last two $p$-factors in (4.12) together with the multiplicity 
vectors as contributions to the abelian vertex operators from which they 
originate (cf.\ eq.\ (3.4)). In this form,
$$ \eqalign{A_+ & =  a \odot E_+P_0 - b^+ \odot E_+P_2 \cr
A_+^+ & =  i b \odot E_+P_1 + a^+ \odot E_+P_3 \cr
B_+ & =  - a^+ \odot E_+P_1 +i b \odot E_+P_3 \cr
B_+^+ & =  -i b^+ \odot E_+P_0 +i a \odot E_+P_2 \cr } 
\eqno(4.13) $$
where $E_+ \equiv E_{+\frac12}$ and $P_i$ project onto the subspaces (3.6). 
We observe that each of the three sectors between which the elementary fields 
in these expressions interpolate arises in two orthogonal copies, 
distinguished by different abelian charge projections in the second $p$-factor.
Again, the $-$ components are determined by (4.6). 

One might proceed and compute also the primary isospin 1 triplet of real 
fermions, e.g., from the operator product expansion of the fields 
(4.12). Inserting the o.p.e.'s for the elementary fields, it is evident
that one will obtain the three permutations of $\psi\odot\eins\odot\eins$
where $\psi$ is the real Ising chiral fermion, along with some Klein
transformations which make the three components anti-commute. Since
nothing exciting happens, we do not present the explicit calculations.

\vskip5mm {\bf 4.3. The 1+1-dimensional Wess-Zumino-Witten (WZW) field at 
level 2} \vskip2mm

The 1+1-dimensional fundamental field of the WZW model is a $2\times 2$ matrix 
with $\SU_L \times \SU_R$ symmetry (by left and right 
matrix multiplication).
It is a local field which carries the quantum numbers of the non-local
primary exchange doublet with respect to both chiral current algebras.

At level 2, it can be represented in the form \cite\Spt 
$$ g_{ij} = (A_i \times A_k + A_i^+ \times A_k^+ +B_i \times B_k + B_i^+ 
\times B_k^+) \; \eps_{kj} \eqno (4.14) $$
where the $\eps$ is necessary because the right action of SU(2) is in the 
conjugate representation. 
This form parallels the representation (3.1a) of the order parameter on its 
own cyclic subspace (we do not consider the analogue of the disorder 
parameter, which would enlarge the joint cyclic Hilbert space and require a 
second copy of $\kk_{\frac12\frac12}$ as in eqs.\ (3.1b).). Due to eq.\ (4.2), 
one has
$$ g_{ij}^* = \eps_{ik}\eps_{jl} \; g_{kl}. \eqno(4.15) $$

We insert the expressions (4.13) for the primary exchange fields into the 
definition (4.14). 
As before, we first consider the interpolation scheme within the joint cyclic
subspace. Every component $g_{ij}$ is a sum of contributions of the form
$X_{ij}^{kl} \odot (E \times E)P_{kl}$ where $(E \times E)P_{kl}$ is the 
$\times$-product of the abelian vertex operators from eq.\ (4.13) on the 
charge sector $H_k \otimes H_l$ according to eq.\ (3.7), the accompanying 
$\times$-product of the elementary fields from eq.\ (4.13) being called 
$X_{ij}^{kl}$. As it turns out, on the cyclic Hilbert space every such 
operator $X_{++}^{kl}$ (resp.\ $X_{+-}^{kl}$) can be identified with a 
contribution to $\sig$ (resp.\ $\mu$) as in eqs.\ (3.1), restricted to a 
$\Gamma$-charge 
sector $H_i$ according to eq.\ (3.6). E.g., the component $g_{+-}$ involves 
the combination $X_{00} = a\times a - b^+ \times b^+$ which appears in the 
representation (3.1a) of $\mu$ on $H_0$. For this identification, no 
multiplicity vectors $u$, $v$ are needed to keep the two copies $H_1$ and 
$H_3$ of $\hh_{\frac1{16}\frac1{16}}$ apart, since the accompanying abelian 
charge projection operators $P_{kl}$ already have the same effect.

The result of these considerations is the following interpolation diagram in 
which the horizontal arrows represent the action of the component $g_{+-}$ 
which involves $\mu$ and $E_{+\frac12} \times E_{+\frac12}$ while the diagonal 
arrows represent $g_{++}$ which involves $\sig$ and $E_{+\frac12} \times 
E_{-\frac12}$ (the other two matrix elements interpolate according to the 
reversed arrows according to eq.\ (4.15)):
$$ \matrix 0 \odot 00 & \longrightarrow & 3 \odot 11 & \longrightarrow & 
0 \odot 22 & \longrightarrow & 3 \odot 33 & \longrightarrow & 0 \odot 00 \\
& \eastcross & & \eastcross & & \eastcross & & \eastcross & \\ 2 \odot 02 & 
\longrightarrow & 1 \odot 13 & \longrightarrow & 2 \odot 20 & \longrightarrow 
& 1 \odot 31 & \longrightarrow & 2 \odot 02 \endmatrix \eqno(4.16) $$
The vertices $i \odot kl$ of this diagram are labelled by the $\Gamma$-charge 
sectors $H_i$ between which $\sig$ and $\mu$ interpolate, and the 
abelian charge sectors $H_k \otimes H_l$ between which the vertex operators 
$E_{\pm\frac12}$ interpolate, according to eqs.\ (3.7). 

The explicit computation of the coefficients yields
$$ \eqalign{ g_{++} \cong \sig & \odot (E_{+\frac12} \times E_{-\frac12})
(-P_{00} + iP_{11} - iP_{22} - P_{33} + iP_{02} - iP_{13} + P_{20} + P_{31})\cr
g_{+-} \cong \mu & \odot (E_{+\frac12} \times E_{+\frac12})
(P_{00} + P_{11} - P_{22} + P_{33} + iP_{02} + P_{13} - iP_{20} - P_{31}) .\cr}
\eqno(4.17) $$
The charge dependences in eqs.\ (4.17) can be removed to a large extent, 
but not entirely, by a unitary similarity transformation of the form
$\eins\odot\sum_{ij} \omega_{ij} P_{ij}$ with suitable complex phases 
$\omega_{ij}$ on the charge sectors $H_i \otimes H_j$. For later use, we 
display two convenient simplifications of eqs.\ (4.17):
$$ \eqalign{g_{++} \cong \sig & \odot (E_{+\frac12} \times E_{-\frac12}) 
\qquad\;\;\cr
g_{+-} \cong \mu & \odot (E_{+\frac12} \times E_{+\frac12}\eu^{-i\pi Q}) 
\qquad\;\;\cr} \eqno(4.18a) $$
or
$$ \eqalign{g_{++} \cong \sig & \odot (E_{+\frac12}\eu^{-i\pi Q} 
\times E_{-\frac12}\eu^{-i\pi Q} ) \cr
g_{+-} \cong \mu & \odot (E_{+\frac12}\eu^{-i\pi Q} \times 
E_{+\frac12}) \cr} \eqno(4.18b) $$
The other matrix elements are given by eq.\ (4.15), i.e., $g_{--} = g_{++}^*$
and $g_{-+} = - g_{+-}^*$.

It is amusing to observe how the Klein transformations $\eu^{-i\pi Q}$ 
in these expressions ``mimic'' the same dual commutation relations
for the abelian vertex operator fields as for the order and disorder fields
(eq.\ (3.3)), and hence serve to cancel the corresponding signs such that the 
$p$-products are indeed local. It is also possible to transfer the Klein 
transformation in eq.\ (4.18) from the vertex operators to the order and 
disorder fields, changing their dual commutation relations into anyonic ones.

In the representation (4.18a), the trace field $\hbox{Tr}\, g$ looks
particularly simple, and with eq.\ (1.7a) turns into
$$ \hbox{Tr}\, g \cong \sqrt2 \; (\sig\odot\mu\odot\mu)
\cong \sqrt2 \; \sig\odot\sig\odot\sig . \eqno(4.19) $$
The latter equivalence holds since the cosine $\mu\odot\mu$ differs from the 
sine $\sig\odot\sig$ by a U(1) transformation, cf.\ Sect.\ 2.1.

On the line $x=0$, the fields $g_{ij}$ can be represented as abelian vertex 
operators, e.g., inserting (1.2a) into (4.18a)
$$ \eqalign{g_{++}\vert_{x=0} & \cong E_\b (P_0+P_2) + E_\gamma (P_1+P_3) \cr
g_{+-}\vert_{x=0} & \cong E_\delta (P_0-P_2) +iE_\eps(P_1-P_3) \cr} 
\eqno(4.20) $$
where $\b = \frac12(1,1,-1)$, $\gamma = \frac12(-1,1,-1)$, $\delta = \frac12
(-1,1,1)$, $\eps = \frac12(1,1,1)$ (with $\b+\delta = \gamma+\eps$), and $P_i$
project on the subspaces of charge $(\gamma-\delta)Q = (i \mod 4)\frac12$, or 
equivalently on the subspaces of right-handed WZW isospin component $j_R^3 = 
(i \mod 4)\frac12$. These formulae are another example of ``abelianization'' 
of correlations along the line, just like Kadanoff's result for the Ising 
model. Note, however, the non-trivial angles between the charge vectors.

It is interesting to verify the unitarity of the matrix-valued field $g_{ij}$ 
in these representations, and to compute the associated chiral SU(2) currents.
Of course, one has to consider operator quadratic expressions of the form
$g_{ij}g_{kj}^*$ or $g_{ij}\partial g_{kj}^*$ which have to be 
renormalized. For conformal fields of chiral scaling dimensions $(h,\bar h)$ 
the proper way of doing so is a point-split regularization and renormalization 
by $\Delta_L^{-2h}\Delta_R^{-2\bar h}$ in order to single out the leading 
contribution to the operator product expansion. This indeed reproduces 
unitarity (using $\Double{\sig\sig} = \Double{\mu\mu} = 
\Double{E_{\pm\frac12} E_{\mp\frac12}} = \eins$; actually one finds that 
rather $g/\sqrt2$ is 
unitary since we have normalized every matrix element such that its two-point 
function $\vac{g_{ij}^*g_{ij}}$ is $(\Delta_L\Delta_R)^{\frac38}$), and gives 
the level 2 currents $J_L^a = i\hbox{Tr}\,\tau^a \Double{\partial_+g g^*} 
\cong J^a \times \eins$ in the form
$$ \eqalign{J^3 & \cong \qquad\qquad \;-\eins \odot 2\partial\varphi \, 
\cong (\a_1 + \a_2)\partial\varphi \cr
J^+ & \cong (E_{+1}+E_{-1}) \odot E_{-1} \cong E_{\a_1} + E_{\a_2} , \cr} 
\eqno(4.21) $$
where $\a_1 = (1,-1)$, $\a_2 = (-1,-1)$, and similar for the right-moving 
currents $J_R^a = i\hbox{Tr}\,\tau^a \Double{\partial_- g^*g} 
\cong \eins \times J^a$. In order to obtain this result, we have used the 
relevant leading contributions to the operator product expansions 
$E_{\pm\frac12}(x)E_{\pm\frac12}(y) \approx \Delta(x-y)^{-\frac14} E_{\pm 1} 
+ \cdots$ and $E_{\pm\frac12}E_{\mp\frac12} \approx \Delta^{\frac14} 
(\eins \pm\frac i2(x-y) \partial\varphi + \cdots)$,
as well as $\sig\sig \approx \mu\mu \approx (\Delta_L\Delta_R)^{\frac18} 
\eins + \cdots$ and $\mu\sig \approx i\sig\mu \approx \sqrt{i}
\Delta_L^{-\frac38}\Delta_R^{\frac18}\psi_L + \cdots $ where the left-moving 
real chiral Ising fermion $\psi_L$ can in turn be represented in the form 
$\frac12(E_{+1}+E_{-1}) \times \eins$, up to another Klein transformation 
which makes $\psi_L$ and $\psi_R$ anti-commute. The form 
(4.21) complies with eq.\ (2.14) with two Frenkel-Kac representations at 
level 1, if one changes the basis to $\a_1 = (\sqrt2,0)$, $\a_2 = (0,\sqrt2)$.

\vskip5mm {\bf 4.4. The determinant field} \vskip2mm 

Every $p$-quadratic expression in either the diagonal or the off-diagonal 
matrix elements $g_{ij}$, eqs.\ (4.18), can be expressed in terms of abelian 
vertex operators alone, since the doubled non-abelian factors $\mu$ and 
$\sig$ can. This observation generalizes the Schroer-Truong formulae (1.7).

A particularly simple representation of the sine form (1.11) can be obtained 
for the $p$-determinant field (which is invariant under $\SU_L \times \SU_R$)
$$ \hbox{Det}_\odot\, g := \textstyle \frac12\;\eps_{ik}\eps_{jl} \;
g_{ij} \odot g_{kl} = \frac12(g_{11} \odot g_{22} + g_{22} \odot g_{11} 
- g_{12} \odot g_{21} - g_{21} \odot g_{12}) \; . \eqno(4.22) $$

In order to compute the $p$-determinant, we take the liberty to use the 
representations (4.18a) and (4.18b) for the first and second $p$-factor, 
respectively.
As for the abelian vertex operators in eqs.\ (4.18), we observe that they
always combine in (4.22) in the form $E_{\pm\frac12}\odot E_{\mp\frac12} 
\cong E_{\pm\a}$ where $\a = \frac1{\sqrt2}$. Consequently, all charges in 
the cyclic Hilbert space are integer multiples of $\a$, and the charge 
operators involved may be replaced by $\eu^{i\pi Q}\odot\eins \cong 
\eu^{i\pi\a Q}$ and $\eins\odot\eu^{i\pi Q} \cong \eu^{-i\pi\a Q} = 
\eu^{i\pi\a Q}\cdot C$ where $C = \eu^{2\pi i\a Q} = C\inv$ is the Casimir 
operator encountered previously in Sect.\ 2.3. 

The crucial point now is that an enhanced symmetry emerges on the
vertex operators of charge $\pm\a$ which was not present on the vertex 
operators of charge $\pm\frac12$ in the representation (4.13), i.e., before 
taking the $p$-square (and which spoils the original 
(squared) level 2 affine chiral SU(2) symmetries generated by (4.21)). 
Namely, all terms contributing to (4.22) 
involve $E_{\pm\a}\eu^{i\pi\a Q} \times E_{\pm\a}\eu^{i\pi\a Q}$, i.e., the 
components of the pseudoreal SU(2) level 1 doublet $\phi_\pm$ as in eq.\ 
(2.13), possibly along with some Casimir operators. Explicitly, we have
$$ \eqalign{ \hbox{Det}_\odot\, g & \cong \frac12 \big[ \sig_D 
\odot (\phi_+C \times \phi_-C + \phi_- \times \phi_+) - i \mu_D
\odot (\phi_+C \times \phi_+C + \phi_- \times \phi_-) \big] \cr
& \qquad \cong \frac1{2\sqrt2i} \left[ (E_{-\frac12} \times E_{+\frac12}) 
\odot (\phi_+ + \phi_-) \times (\phi_+ + \phi_-) \right. \cr 
& \qquad\qquad\qquad\qquad \left. +(E_{+\frac12} \times E_{-\frac12}) 
\odot (\phi_+ - \phi_-) \times (\phi_+ - \phi_-) \right] . \cr } 
\eqno(4.23) $$
For the second line, we observed that on the cyclic Hilbert space
$C \otimes C$ equals $\eins$, and we evaluated $\sig_D$ and $\mu_D$ according 
to eqs.\ (1.7).  

The new level 1 chiral SU(2) symmetries permit to rotate the components
$(\phi_+ \pm \phi_-)$ into $\sqrt2 \phi_\mp$ in the left chiral factor and 
into $\pm\sqrt2 \phi_\pm$ in the right chiral factor. This unitary similarity 
transformation turns the determinant field into 
$$ \hbox{Det}_\odot\, g \cong \frac1{\sqrt2i} \left[ (E_{-\frac12} \times 
E_{+\frac12}) \odot (\phi_- \times \phi_+) - (E_{+\frac12} \times 
E_{-\frac12}) \odot (\phi_+ \times \phi_-) \right] . \eqno(4.24) $$
After this transformation, the cyclic Hilbert space is seen to carry opposite 
charges in the two chiral factors. We may therefore insert $\eins = 
\eins\odot(\eu^{i\pi\a Q} \otimes \eu^{i\pi\a Q})$ and re-substitute the 
vertex operators $E_{\pm\a}$ for the doublet $\phi_\pm$. This gives finally
$$ \hbox{Det}_\odot\, g \cong \frac1{\sqrt2i}(E_{-\b} \times E_\b - 
E_\b \times E_{-\b}) \eqno(4.25) $$ 
with $\b = \frac12(1,\sqrt2)$ or, after an O(2) rotation in charge space, 
$\b = \frac12 \sqrt3$. The non-triviality of this formula can be esteemed 
upon the intricate exercise to test it only at the four-point level.

\vskip5mm {\bbf 5. Conclusions} \vskip3mm

We have established a sort of ``Quantum Field Lego'' with chiral non-abelian
exchange fields which can be recombined in a large variety of ways to produce
several local and non-local, chiral and 1+1-dimensional Wightman fields 
belonging to different models. 
It is the operator formulation of some remarkable findings in 
the critical Ising model by Kadanoff and Ceva and by Schroer and Truong, and 
predicts, among other things, generalizations thereof in the context of a WZW
(= generalized Thirring) model.

A prominent feature of the basic decomposition (3.4) is that it produces
non-abelian exchange fields, satisfying commutation relations according
to a non-abelian representation of the braid group and possessing
a non-abelian fusion structure, by factorization of abelian vertex operators
with anyonic commutation relations and additive charge structure.

We have seen along the way (Sects.\ 2.3 and 4.3) how the generators of the 
chiral symmetry can be explicitly computed in our operator representations
and are found, of course, in complete agreement with well-known results.

Throughout the paper, we have mentioned only en passant the issue of braid 
group 
commutation relations. Let us just add that $p$-products go along with the 
tensor products of representations of the braid group; notably the basic 
result singles out an abelian subrepresentation within the square of the 
Hecke type representations associated with the elementary fields.

Let us consider the question whether one may expect parallel findings 
in other minimal or coset models. Since abelian vertex operators are the 
primary fields for stress-energy tensors with integer central charge $c$, 
one may expect representations of abelian vertex operators as $p$-products 
of non-abelian ones whenever a stress-energy tensor with integer central 
charge occurs in the product of the chiral observables. In the case of the 
doubled Ising model, the latter is just the sum of the two stress-energy 
tensors with $c=\frac12$, and in the case of the determinant field, it is the 
coset stress-energy tensor with $c=1$ for the coupling of two level 2 current 
algebras into the level 4 current algebra. Admitting sufficiently high 
$p$-powers, any model with rational central charge presumably provides 
another set of non-abelian roots 
of abelian vertex operators. On the other hand, the ``Kadanoff mechanism'' of 
quadratic abelianization upon restriction to a time-like axis, which in the 
present model we consider to be the reason for the possibility of several 
independent recombinations of the same elementary fields, can only work when 
the two opposite chiral central charges add up to an integer. This
property of course distinguishes the Ising model with $c=\frac12$ and the 
SU(2) model at level 2 with $c=\frac32$ studied in this paper from 
other minimal or WZW models.

An immediate further application, not treated in this note, is the 
construction of the primary field $\phi_{21}$ with dimension $\frac7{16}$ of 
the minimal 
model with $c=\frac7{10}$ which arises in the coset construction as the 
singlet contribution to the $p$-product of the SU(2) primary doublets at 
level 1 (eq.\ (2.13)) and at level 2 (eq.\ (4.12)), commuting with the level 3
currents.

Although the basic result involves chiral and non-local fields, several of 
our applications pertain to ordinary local Wightman fields in 1+1-dimensional
Minkowski space-time.

We want to draw the attention to a point of general importance which is 
illustrated by eq.\ (4.25). By definition, the determinant field
is $p$-quadratic in the four mutually local fields $g_{ij}$.
On the other hand, the right hand side of eq.\ (4.25), by elementary 
identities using the representation (1.7), equals one half of
$$ \sig_D \odot \sig_D \odot \sig_D - \sig_D \odot \mu_D \odot \mu_D 
- \mu_D \odot \sig_D \odot \mu_D - \mu_D \odot \mu_D \odot \sig_D \; .
\eqno(5.1) $$ 
(This formula would not be guessed from eqs.\ (4.18) according to which 
$g_{ij}$ are not $p$-cubic in $\sig$ and $\mu$.) The determinant field is 
thus at the same time $p$-cubic in local fields.  
Yet, a common sixth-order factorization into local fields cannot be perceived 
(at least not in terms of our elementary fields). This indicates, with 
due precaution, that something like a ``prime decomposition'' with respect to 
the $p$-product does not exist within the class of {\it local} Wightman fields.

We consider our work as an exercise with non-canonical fields from which
several lessons on the general theory of Wightman fields can be drawn.
We want to refute the possible impression that the construction of 
$p$-products is just an artificial trick to construct new Wightman fields 
without a proper physical meaning from old ones. On the contrary, all the 
product fields in this article were previously known and considered in their 
own model context. The new aspect is the passage between different models
enabled by $p$-products.

\bgap  {\bf Acknowledgment:} 
\hangafter=1 \hangindent 33mm 
The second author (K.-H.R.) thanks B. Schroer for many helpful and 
interesting comments during the final preparation of this article.

\newpage
\bgap {\bbf References} \vskip2mm

\def\ref#1{\par \noindent \hangafter=1 \hangindent 14pt \cite{#1}}
\parskip 3.1pt
\baselineskip=2.5ex
\def\CMP#1{Com\-mun.\ Math.\ Phys.\ {\bf #1}}
\def\LMP#1{Lett.\ Math.\ Phys.\ {\bf #1}}
\def\PR#1{Phys.\ Rev.\ {\bf #1}}
\def\PL#1{Phys.\ Lett.\ {\bf #1}}
\def\NP#1{Nucl.\ Phys.\  {\bf #1}}
\ref\K\ L.P. Kadanoff: {\it Correlations along a line in the two-dimensional
Ising model}, \PR{188} (1969) 859--863; \newline
H. Ceva, L.P. Kadanoff: {\it Determination of an operator algebra 
for the two-dimensional Ising model}, \PR{B 3} (1971) 3918--3939.
\ref\ST\ B. Schroer, T.T. Truong: {\it The order/disorder quantum field 
operators associated with the two-dimensional Ising model in the continuum
limit}, \NP{B 144} (1978) 80--122.
\ref\W\ A.S. Wightman: {\it Introduction to some aspects of the relativistic 
dynamics of quantized fields}, in: M. Levy (ed.), Carg\`ese Lectures in
Theoretical Physics 1964; New York: Gordon and Breach, 1967, pp.\ 171--291. 
\ref\B\ H.-J. Borchers: {\it Field operators as $C^\infty$ functions in 
spacelike directions}, Nuovo Cim.\ {\bf 33} (1964) 1600--1613.
\ref\RS\ K.-H. Rehren, B. Schroer: {\it Exchange algebra on the light-cone
and order/disorder $2n$-point functions in the Ising field theory},
\PL{B 198} (1987) 84--88.
\ref\BPZ\ A.A. Belavin, A.M. Polyakov, A.B. Zamolodchikov: {\it Infinite
dimensional symmetries in two-dimensional quantum field theory},
\NP{B 241} (1984) 333--380.
\ref\GKO\ P. Goddard, A. Kent, D. Olive: {\it Unitary representations of the 
Virasoro and super-Virasoro algebras}, \CMP{103} (1986) 105--119.
\ref\BRSS\ L.V. Belvedere, K.D. Rothe, B. Schroer, J.A. Swieca: {\it
Generalized two-dimensional abelian gauge theories and confinement},
\NP{B 153} (1979) 112--140.
\ref\KZ\ V.G. Knizhnik, A.B. Zamolodchikov: {\it Current algebra and 
Wess-Zumino model in two dimensions}, \NP{B 247} (1984) 83--103.
\def\ref#1{\par \noindent \hangafter=1 \hangindent 19pt \cite{#1}}
\ref\WZW\ E. Witten: {\it Non-abelian bosonization in two dimensions},
\CMP{92} (1984) 455--472.
\ref\F\ K. Frieler: Diploma thesis, Hamburg University 1997 (in german).
\ref\LS\ H. Lehmann, J. Stehr: {\it The Bose field structure associated with 
a free massive Dirac field in one space dimension}, DESY preprint 76-29 (1976).
\ref\SS\ B. Schroer, J.A. Swieca: {\it Conformal transformations for 
quantized fields}, \PR{D 10} (1974) 480--485.
\ref\CRW\ A.L. Carey, S.N.M. Ruijsenaars, J.D. Wright: {\it The massless 
Thirring model: positivity of Klaiber's $n$-point functions}, \CMP{99} 
(1985) 347--364.
\ref\M\ S. Mandelstam: {\it Soliton operators for the quantized sine-Gordon
equation}, \PR{D 11} (1975) 3026--3030.
\ref\BB\ K.-H. Rehren: {\it Bounded Bose fields}, to appear in \LMP{}
\ref\SSV\ B. Schroer, J.A. Swieca, A.H. V\"olkel: {\it Global operator 
expansions in conformally invariant relativistic quantum field theory},
\PR{D 11} (1975) 1509--1520.
\ref\FRS\ K. Fredenhagen, K.-H. Rehren, B. Schroer: {\it Superselection 
sectors with braid group statistics and exchange algebras, II},
Rev.\ Math.\ Phys.\ (Special Issue 1992) 113--157.
\ref\J\ M. J\"or\ss: {\it The construction of pointlike localized charged 
fields from conformal Haag-Kastler nets}, \LMP{38} (1996) 257--274.
\ref\VO\ I.B. Frenkel, J. Lepowsky, A. Meurman: {\rm Vertex Operator Algebras 
and the Monster}, Pure and Appl.\ Math.\ {\bf Vol.\ 134}; Boston: Acad.\ 
Press, 1988; especially Chap.\ 8.
\ref\KF\ I.B. Frenkel, V.G. Kac: {\it Basic representations of affine Lie 
algebras and dual resonance models}, Invent.\ Math.\ {\bf 62} (1980) 23-66.
\ref\Spt\ K.-H. Rehren: {\it Space-time fields and exchange fields}, 
\CMP{132} (1990) 461--483.
\bye